\documentclass[twocolumn,secnumarabic,amssymb, nobibnotes, aps, prd]{revtex4-1}

\usepackage{graphicx}
\usepackage{hyperref}
\usepackage{amsmath}
\usepackage{xcolor}
\begin{document}

\title{Wormholes in $f(R,T)$ gravity satisfying the null energy condition with isotropic pressure}

\author{Ayan Banerjee}
\email{ayan\_7575@yahoo.co.in}
\affiliation{Astrophysics and Cosmology Research Unit, University of KwaZulu Natal, Private Bag X54001, Durban 4000,
South Africa.}

 \author{M. K. Jasim}
 \email{mahmoodkhalid@unizwa.edu.om}
\affiliation {Department of Mathematical and Physical Sciences, College of Arts and Science, University of Nizwa, Nizwa, Sultanate of Oman}

 \author{Sushant G. Ghosh}
 \email{sgghosh@gmail.com, sghosh2@jmi.ac.in}
\affiliation {Centre for Theoretical Physics, Jamia Millia Islamia, New Delhi 110025, India}
\affiliation{Astrophysics and Cosmology Research Unit, University of KwaZulu Natal, Private Bag X54001, Durban 4000,
South Africa.}

\begin{abstract}
We consider the $f(R, T)$ theory of gravity, in which the gravitational Lagrangian is given by an arbitrary function of the Ricci scalar and the trace of the energy-momentum tensor, to study static spherically symmetric wormhole geometries sustained by matter sources with isotropic pressure. According to restrictions on the wormhole geometries, we carefully adopt different strategies to construct solutions with the properties and characteristics of wormholes. Using an utterly general procedure, we provide several examples of wormholes in which the matter threading the wormhole throat satisfies all of the energy conditions and discuss general mechanisms for finding them. Finally, we postulate a smooth transformation for simplifying the nonlinear field equations and have more consistent results than the other ones to conclude that the results can be viewed as specific exact wormhole solutions without exotic matter.
\end{abstract}

\keywords{$f(R,T)$ gravity; Isotropic Pressure; Wormhole Solution}

\pacs{04.20.Gz, 11.27.+d, 04.62.+v, 04.20.−q}
\maketitle

\section{Introduction}\label{sec1}
Wormholes act as tunnels from one region of
spacetime to another, perhaps through which observers can freely traverse. 
Historically, Flamm in 1916 \cite{flam} showed Schwarzschild solution represents a wormhole. Sailing through history, one finds, in the middle of the 1935s,  Einstein and Rosen \cite{EinsteinRosen} implicitly showed a
a curved-space structure can join two  distinct regions of spacetime through
a tunnel-like curved spatial, namely `Einstein-Rosen bridge' (ERB).
However, their solution was shown to be an invalid particle model as the mass-energy of such a curved-space topology is the order of 
Planck mass. The field laid dormant until Wheeler revived the subject in the 1950s. He was interested in topological issues in general relativity (GR), which was denoted ``gravitational-electromagnetic entity" in short geons, and he also coined the term wormhole.  But Wheeler and Fuller \cite{Fuller1962} wormholes would collapse instantly upon formation. In fact, ERB is a non-traversable wormhole, even by a photon that was transformed later into Euclidean wormholes by Hawking \cite{Hawking} and others \cite{Betzios:2019rds,ArkaniHamed:2007js,Hebecker:2018ofv}.

Modern interest in wormholes is rejuvenated mainly because of the classical paper by Morris and Thorne \cite{Morris:1988cz}.  Morris, Thorne and Yurtsever \cite{Morris:1988tu} came up with other possibilities of constructing time machines that violate Hawking's chronology protection conjecture \cite{Hawking1992}. Such wormholes possess a peculiar property, namely exotic matter, to maintain the wormhole tunnel open. Roughly speaking, the matter violates the weak/null energy conditions called `exotic matter' \cite{Visser} at least in a neighbourhood of the wormhole throat. Such strange object exists both in the static \cite{Anabalon:2012tu,Balakin:2010ar,Jamil:2009vn,Cataldo:2002jw} as well as dynamic \cite{Dehghani:2009xu,Bochicchio:2010df,DeBenedictis:2008qm,GonzalezDiaz:2003pb,Cataldo:2011zn,Hansen:2009kn} cases, and sustained by a single fluid component. Thus, minimize the use of exotic matter is always a subject of an intense research area. However, exotic matter sounds unusual in classical relativity, but quantum field theory appears to violate this condition as a natural consequence of the topology of spacetime that fluctuates in time. Such an example is the standard Casimir system of parallel plates, for which there is a negative energy density between the plates. With the development of cut-off of the stress-energy, by matching the interior solution of metric to an exterior vacuum spacetime, 
at a junction interface, one may confine the exotic matter at the throat of the wormhole \cite{Visser:1989kh,Visser:1989kg}.

From the above discussion, we can say that NEC is an artefact of GR to sustain a wormhole. However, wormholes in the context of modified gravity it is shown that matter sources that do not violate the NEC as the higher-order curvature terms that support these exotic geometries. It is illustrated that modified theories of gravity could explain the
late-time accelerated expansion of the Universe (arXiv:0807.1640 [gr-qc]). Consequently, a wide variety of wormhole solutions have been analyzed in various modified theories of gravity. More recently, Lobo and Oliveira \cite{Lobo:2009ip} showed that there exists a wormhole solution satisfying all of the energy conditions in $f(R)$ gravity at wormhole throat and several generalizations of this solution can be found in Refs. \cite{Mazharimousavi:2012xv,Pavlovic:2014gba,Sharif:2018jdj,DeBenedictis:2012qz}. In the curvature-matter coupled generalization of $f(R)$ gravity, wormhole solutions were also analyzed \cite{MontelongoGarcia:2010xd} and hybrid metric-Palatini types of $f(R)$ gravity \cite{Rosa:2018jwp}.

On the other hand, modified theories have the freedom to play with the effective stress-energy tensor so that one can find an alternate route to solve the exotic matter problem and it respects all the energy conditions.
One of the modified gravity theories that attracted significant attention is $f(R,T)$ gravity where the Lagrangian is an arbitrary function 
of Ricci scalar $R$ and the trace of the energy-momentum tensor $T$. This theory was developed by
Harko \emph{et al} \cite{Harko:2011kv}. They derived the gravitational field equations in the metric formalism and the equations of motion of test particles, which follow from the covariant divergence of the stress-energy tensor. Interestingly, the dependence on $T$ may be a consequence of the universe being partially filled by an exotic imperfect fluid or a consequence of quantum effects coming from a conformal anomaly. Several works use the $f(R,T)$ gravity model to solve various cosmological problems and satisfactorily explanation of the late time acceleration of the universe (see details  
\cite{Houndjo:2011fb,jamil2011re,Jamil:2012pf,Baffou:2013dpa,Singh:2013bpa,Sharif:2014ioa,Baffou:2017pao,Mishra:2017sdq}). 

In $f(R,T)$ gravity, wormhole solutions were investigated in \cite{Azizi:2012yv}. These wormholes were found to satisfy the NEC in some subspaces depending on the model parameter. It is also noticeable that $f(R,T)$ wormholes are also sensitive depending on various types of form functions, such as  $f(R,T) =R+\alpha R^2+\lambda T$ \cite{Sharma:2021jht}. The wormhole solutions  have been studied depending on the various choice of coupling parameters $\alpha$ and $\lambda$, and showed that the violation of energy conditions are also avoidable \cite{Sharma:2021jht}. In this proceedings, wormhole solutions have also investigated in \cite{Zubair:2016cde,Moraes:2016akv,Elizalde:2018frj,Elizalde:2018arz,Banerjee:2019wjj,Shweta:2020fxs} with different physical 
inputs. Motivated by the article \cite{Cataldo:2016dxq}, we aim to obtain the wormhole solutions with a single perfect fluid, i.e. a matter source with isotropic pressure. More specifically, we will specify the shape and redshift functions and establish the possible existence of appropriate static wormhole configurations. 

This paper is organized as follows: In the next Sec. \ref{sec2}, we consider the relevant action and write out the gravitational field equations for $f(R,T)$ gravity. The detailed analysis field equations for the Morris-Thorne metric within the framework of $f(R,T)$ gravity is the subject of Sec. \ref{sec3}. Sec. \ref{sec4} is devoted to introducing specific redshift and shape functions to discuss wormhole geometry as well as energy conditions. Next, we transform the isotropic equation to a second-order homogeneous equation and found a wormhole solution considering a specific shape function in \ref{sec5}, which does not violate the NEC. In Sec. \ref{sec6}, we check the equilibrium condition for the obtained
solutions. Finally, we provide concluding remarks in Sec. \ref{sec7}.

We have used units which fix the speed of light and
the gravitational constant via $c=G=1$.


\section{$f(R,T)$ formalism } \label{sec2}
In this section we briefly review  the $f(R, T)$ gravity.   
The starting point is the Einstein-Hilbert  gravitational action  \cite{Harko:2011kv} which can be recast as  
\begin{eqnarray}\label{eq1}
S=\frac{1}{16\pi}\int f(R,T)\sqrt{-g} d^{4}x+\int \mathcal{L}_{m}\sqrt{-g}d^{4}x, \label{eq2}
\end{eqnarray}
where $f(R,T)$ is an arbitrary generic function of $R$ and $T$, the Ricci scalar and the trace of the energy momentum tensor $T_{\mu\nu}$, respectively. Moreover, $g$ is the determinant of the metric $g_{\mu\nu}$, 
and $\mathcal{L}_{m}$ is the matter Lagrangian. The energy-momentum tensor of matter is defined as 
\begin{eqnarray}
T_{\mu\nu}=-\frac{2}{\sqrt{-g}}\,\frac{\delta(\sqrt{-g}\,\mathcal{L}_m)}{\delta g^{\mu\nu}}, \label{eq3}
\end{eqnarray}
with the trace $T=g^{\mu\nu}T_{\mu\nu}$. Here the Lagrangian density $\mathcal{L}_m$ of matter depends only on the metric tensor components $g_{\mu\nu}$, and not its derivative. Thus, we have 
\begin{eqnarray}
T_{\mu\nu}=g_{\mu\nu} \mathcal{L}_m-\frac{2\,\partial(\mathcal{L}_m)}{\partial g^{\mu\nu}}. \label{eq4}
\end{eqnarray}

Varying the action (\ref{eq1}) with respect to the metric tensor $g^{\mu\nu}$, we obtain the field equations of $f(R, T)$ gravity \cite{Harko:2011kv} as
\begin{multline}\label{eq5}
\delta S =\frac{1}{16\pi} \int \Big[ f_{R}(R,T)R_{\mu\nu}\delta  g^{\mu\nu} + f_{R}(R,T)g_{\mu\nu}\Box \delta  g^{\mu\nu}
\\
 -f_{R}(R,T) \nabla_{\mu}\nabla_{\nu} \delta g^{\mu\nu}+f_{T}(R,T)\frac{\delta(g^{\alpha\beta}T_{\alpha\beta})}{\delta g^{\mu\nu}}\delta g^{\mu\nu} \\
 -\frac{1}{2}g_{\mu\nu}f(R,T)\delta g^{\mu\nu}+16 \pi
 \frac{1}{\sqrt{-g}}\frac{\delta(\sqrt{-g}\,\mathcal{L}_m)}{\delta g^{\mu\nu}} \Big]\sqrt{-g}d^{4}x ,
\end{multline}
where $f_R (R,T)={\partial f(R,T)}/{\partial R}$ and $f_T (R,T)={\partial f(R,T)}/{\partial T}$. 
By the variation of $T$ with respect to $g^{\mu\nu}$, we obtain
\begin{eqnarray}\label{eq5a}
\frac{\delta(g^{\alpha\beta}T_{\alpha\beta})}{\delta g^{\mu\nu}}\delta g^{\mu\nu}=T_{\mu\nu}+\Theta_{\mu\nu}.
\end{eqnarray}
We should note that $\nabla_\mu$ is associated with the Levi-Civita connection of metric tensor $g_{\mu\nu}$ and box operator $\Box$ is defined by
\begin{eqnarray}
\Box\equiv\partial_\mu(\sqrt{-g}g^{\mu\nu}\partial_\nu)/\sqrt{-g}, ~~~~ \textrm{and} ~~~~ \Theta_{\mu\nu}=g^{\alpha\beta}\delta T_{\alpha\beta}/ \delta g^{\mu\nu}. \nonumber
\end{eqnarray} 
Now, performing by parts integration to the second and third terms in Eq. (\ref{eq5}), one can obtain the 
following modified field equation as
\begin{eqnarray}\label{eq6}
\left( R_{\mu\nu}- \nabla_{\mu} \nabla_{\nu} \right)f_R (R,T) +\Box f_R (R,T)g_{\mu\nu} - \frac{1}{2} f(R,T)g_{\mu\nu}  \nonumber \\ = 8\pi\,T_{\mu\nu}  - f_T (R,T)\, \left(T_{\mu\nu}  +\Theta_{\mu\nu}\right) \nonumber \\.
\end{eqnarray}
In this approach, if $f (R, T)$ $\equiv f(R)$, then we get back to the field equations for $f(R)$ gravity and when $f (R, T)$ $\equiv R$, standard Einstein's field equations are recovered in GR.

The covariant derivative of the field equations (\ref{eq6}) gives \cite{Houndjo:2011fb}
\begin{eqnarray}\label{eq7}
\nabla^{\mu}T_{\mu\nu}=\frac{f_T(R, T)}{8\pi -f_T(R,T)}\bigg[(T_{\mu\nu}+\Theta_{\mu\nu})\nabla^{\mu}\ln f_T(R,T)\nonumber\\ +\nabla^{\mu}\Theta_{\mu\nu}-\frac{1}{2}g_{\mu\nu}\nabla^{\mu}T\bigg].~~~~~~
\end{eqnarray}
It is important to stress that, in $f(R,T)$ gravity, the stress-energy tensor of the matter fields is not conserved
due to the interaction between the matter and curvature. The Eq. (\ref{eq5a}), on using the 
Eq. (\ref{eq3}), yields:
\begin{equation}\label{eq8}
\Theta_{\mu\nu}= - 2 T_{\mu\nu} +g_{\mu\nu}\mathcal{L}_m - 2g^{\alpha\beta}\,\frac{\partial^2 \mathcal{L}_m}{\partial g^{\mu\nu}\,\partial g^{\alpha\beta}}.~~~
\end{equation}
We assume that the matter is a perfect fluid having the usual stress-energy tensor
\begin{equation}\label{eq9}
T_{\mu\nu}=(\rho+p)u_\mu u_\nu-p g_{\mu\nu},
\end{equation}
where $u^{\mu}$ is the four-velocity such
that $u^{\mu}u_{\mu} = 1$ and $u^\mu\nabla_\nu u_\mu=0$
with $\rho$ and $p$ are, respectively, the energy-density and
 pressure.
 
 For the perfect fluid the matter Lagrangian density can be taken as $\mathcal{L}_m = -p$, which leads to
\begin{equation}\label{eq10}
\Theta_{\mu\nu}=-2T_{\mu\nu}- p g_{\mu\nu}.
\end{equation}

We begin our study of the isotropic wormhole aspects  on the assumption of linear functional form $f(R,T)=R+2\chi T$ \cite{Harko:2011kv}, where $\chi$ is a constant. With this assumptions, the field Eq. (\ref{eq6}) simplifies to
\begin{equation}\label{eq11}
G_{\mu\nu}=8\pi T_{\mu\nu}+\chi
Tg_{\mu\nu}+2\chi(T_{\mu\nu}+pg_{\mu\nu}),
\end{equation}
where $G_{\mu\nu}$ is the Einstein tensor. The considered specific case has been widely studied in the cosmological context of $f(R,T)$ gravity~\cite{Houndjo:2011fb}. Note that
when $\chi=0$, one can easily recover the GR case. Moreover, substituting $f(R,T)= R+2\chi T$, in Eq. (\ref{eq7}) leads to the relation
\begin{equation}\label{eq12}
(8\pi+2\chi)\nabla^{\mu}T_{\mu\nu}=-2\chi\left[\nabla^{\mu}(pg_{\mu\nu})+\frac{1}{2}g_{\mu\nu}\nabla^{\mu}T\right].
\end{equation}
However, one should add a cautionary note about curvature-matter coupling corresponding to a non-vanishing covariant derivative of the energy-momentum tensor. In this sense, one may interpret the curvature-matter coupling as an exchange of energy and momentum between both. However, this type of theory leads to non-geodesic motion and, consequently, leads to the appearance of an extra force \cite{Bertolami:2007gv}. In fact, the extra force is always orthogonal to the four-velocity of the massive particles moving in the gravitational field, and the corresponding acceleration law was obtained in the weak field limit 
\cite{Capozziello:2014bqa,Harko:2020ibn,Harko:2008qz}.

\section{The wormhole geometry and the field equations} \label{sec3}
The general static spherically symmetric line element representing a wormhole is given by \cite{Morris:1988cz}
\begin{equation}\label{eq13}
ds^2=-e^{2\Phi(r)}dt^2+\frac{dr^2}{1-\frac{b(r)}{r}}+r^2(d\theta^2+\sin^2\theta d\phi^2),
\end{equation}
where $\Phi(r)$ and $b(r)$, respectively, are the redshift and the shape functions. Note that the non-monotonic radial coordinate $r$ decreases from infinity to a minimum surface radius $b(r_0)=r_0$, where the throat of the wormhole is located, and then it increases
from $r_0$ back to infinity. $\Phi(r)$ must be finite everywhere to avoid the presence of an event horizon,
which are identified as the surfaces with $e^{2\Phi(r)}\rightarrow 0$.

The shape function $b(r)$ must satisfy the so-called \textit{flaring-out condition} given by  $\frac{b(r)-rb^{\prime}(r)}{b^2(r)}>0$ \cite{Morris:1988cz}, which reduces to become $b^{\prime}(r_0) < 1$ at the throat $r = r_0$. Another condition
that needs to be satisfied is $1- b(r)/r > 0$.
Here and hereafter the prime denotes the derivative with respect to the radial coordinate $r$.
On the other hand, finiteness of the proper radial distance $\ell(r)$ is given by
\begin{equation}\label{eq14}
\ell (r) = \pm \int^{r}_{r_0}{\frac{dr}{\sqrt{1-\frac{b(r)}{r}}}}  ,
\end{equation}
which is required to be finite everywhere. In fact, the proper distance `$\ell$' should be 
greater than or equal to the coordinate distance, i.e. $ \mid \ell (r) \mid$ $\geq$ $r-r_0$, where the  $\pm$ represents the upper and lower space-time connected by the throat of the wormhole $\ell = 0$.

Now using Eqs. (\ref{eq9}), (\ref{eq11}) and (\ref{eq13}) we have the explicit form of the Einstein field equations in $f(R,T)$ gravity yield \cite{Banerjee:2019wjj}
\begin{eqnarray}
&& \frac{b^{\prime}}{r^{2}} = \left(8\pi+3\chi\right)\rho
-\chi p,\label{eq19}\\
&& 2\left(1-\frac{b}{r}\right)\frac{\Phi^{\prime}}{r}-\frac{b}{r^{3}} = \left(8\pi+3\chi\right)p
-\chi \rho,\label{eq20}\\
&& \left(1-\frac{b}{r}\right)\left[\Phi^{\prime\prime}+\Phi^{\prime 2}-\frac{b^{\prime}r-b}{2r(r-b)}\Phi^{\prime}-\frac{b^{\prime}r-b}{2r^2(r-b)} \right.  \nonumber\\
&& \left. +\frac{\Phi^{\prime}}{r}\right] = \left(8\pi+3\chi\right)p
-\chi \rho. \label{eq21}
\end{eqnarray}

In fact, we can rewrite  the Eqs. (\ref{eq19}) and (\ref{eq20}) in terms of energy density ($\rho$) and 
pressure isotropy ($p$) equation, which are

\begin{eqnarray}
 \rho &=& \frac{r \left((3 \chi +8 \pi ) b'+2 r \chi  \Phi '\right)-b \left(2 r \chi  \Phi '+\chi \right)}{8 r^3 \left(\chi ^2+6 \pi  \chi +8 \pi ^2\right)},
\label{sp1} \\ 
p &=& \frac{r \left(\chi  b'+2 r (3 \chi +8 \pi ) \Phi '\right)-(3 \chi +8 \pi ) b \left(2 r \Phi '+1\right)}{8 r^3 \left(\chi ^2+6 \pi  \chi +8 \pi ^2\right)},\label{sp2} \nonumber \\
\end{eqnarray}
while the equation of pressure isotropy reduces to
\begin{equation}
\Phi^{\prime\prime}+ (\Phi^{\prime})^2-\frac{b' r-3b+2r}{2r(r-b)}\Phi^{\prime}
= \frac{b' r-3b}{2r^2(r-b)}. \label{sp3}
\end{equation}
The isotropy equation is the same for the classical general relativity with a perfect fluid source. Therefore any of the well-known metrics found historically will satisfy (\ref{sp3}). Observe that the isotropy equation, which serves as the master differential equation in this analysis containing two unknowns $\Phi(r)$ and $b(r)$. Since we have only three equations, namely, Eqs. (\ref{sp1})-(\ref{sp3}), with four unknown functions of $r$, i.e. $\rho(r)$, $p(r)$,  $\Phi(r)$ and $b(r)$. Obtaining explicit solutions have been found for suitable parameter values through rigorous empirical testing such that the model under investigation satisfies the elementary requirements for physical plausibility.



\section{Some specific solutions}\label{sec4}

In this work we are interested in deducing exact solutions of wormhole in $f(R,T)$ gravity.
Therefore, we adopt the approach followed by 
a physically reasonable form for the shape function or redshift function to close the system of equations.

\subsection{Zero-tidal force wormholes}

Here, we consider a particularly interesting case of constant redshift function, $\Phi' =0$. 
Putting $\Phi' =0$ into the Eq. (\ref{sp3}), we obtain $b(r)=Cr^3$. 
By requiring the condition $b(r_0) =r_0 $, the spacetime metric takes the form
\begin{equation}\label{eq13}
ds^2=- dt^2+\frac{dr^2}{1- (\frac{r}{r_0})^2}+r^2(d\theta^2+\sin^2\theta d\phi^2).
\end{equation}

Taking into account Eqs. (\ref{sp1}) and (\ref{sp2}), the stress-energy tensor profile is given by 
\begin{equation}
\rho=\frac{C (\chi +3 \pi )}{\chi ^2+6 \pi  \chi +8 \pi ^2},~~~~\\
p= -\frac{\pi  C}{\chi ^2+6 \pi  \chi +8 \pi ^2}.
\end{equation}

To verify the flaring-out condition evaluated at the throat we found that $b'(r_0) =3 \nless 1$. This implies that there are no zero-tidal-force wormhole solutions sustained everywhere by an isotropic perfect fluid in $f(R,T)$ gravity.

\subsection{Model with $\Phi (r)=\beta  \log \left(\frac{r_0}{r}\right)$ }

Now, we consider a specific redshift function and taking into account Eq. (\ref{sp3}), leads to the solution of shape function as
\begin{equation}
b(r)= \frac{\beta(2+\beta)}{\beta(2+\beta)-1}r+C_1 r^{\frac{(2\beta-1)(3+\beta)}{\beta-1}}.
\end{equation}

Note that the fundamental condition  $b(r)/r \rightarrow 0$ when $r \rightarrow \infty$ imposes
the restriction on $0.5<\beta< 1$ i.e. the solutions are asymptotically flat. Evaluated at the throat, we have 
\begin{equation}
b(r)= \frac{\beta(2+\beta)}{\beta(2+\beta)-1}r-\tilde{C_1} \left(\frac{r}{r_0}\right)^{\frac{(2\beta-1)(3+\beta)}{\beta-1}},
\end{equation}
where $\tilde{C_1}= \frac{1}{(2\beta+\beta^2-1)} r_0^{-2(2\beta-1+\beta^2/(\beta-1)}$. In order to satisfy the flaring-out condition at the throat, we have
\begin{equation}
b'(r_0)= \frac{\beta  (\beta +2)}{\beta  (\beta +2)-1}-\frac{\tilde{C_1} (\beta +3) (2 \beta -1) }{(\beta -1) r_0},
\end{equation}
the value of $b'(r_0)=0.3 <1$ depending on the value of
$\tilde{C_1}=1 , r_0=1$ and $\beta=0.6$. $\tilde{C_1}$ is always grater than zero with the respective range. The value of $b'(r_0)$ depends on $\tilde{C_1}$ (if we decrease) and $\beta$ goes upto $0.9$ where $b'(r_0)<1$. 

The energy density and pressure are given by the following expressions
\begin{widetext}
\begin{eqnarray}
\rho&=&\frac{(\beta -1) \beta  r ((\beta +3) \chi +4 \pi  (\beta +2))-2 (2 \beta -1) (\beta  (\beta +2)-1) \tilde{C_1} ((\beta +2) \chi +2 \pi  (\beta +3)) \left(\frac{r}{r_0}\right)^{2 \beta +\frac{4}{\beta -1}+7}}{4 (\beta -1) (\beta  (\beta +2)-1)  (\chi +2 \pi ) (\chi +4 \pi )r^3}, \label{m4}\\ 
p &=& -\frac{2 \left(2 \beta ^3+3 \beta ^2-4 \beta +1\right) \tilde{C_1} (\beta  \chi +2 \pi  (\beta -1)) \left(\frac{r}{r_0}\right)^{\frac{(\beta +3) (2 \beta -1)}{\beta -1}}+(\beta -1) \beta  r ((\beta -1) \chi +4 \pi  \beta )}{4 (\beta -1) \left(\beta ^2+2 \beta -1\right)  (\chi +2 \pi ) (\chi +4 \pi )r^3},\label{m5}\\
\rho+p &=&\frac{(\beta -1) \beta  r ((\beta +3) \chi +4 \pi  (\beta +2))-2 (2 \beta -1) (\beta  (\beta +2)-1) \tilde{C_1} ((\beta +2) \chi +2 \pi  (\beta +3)) \left(\frac{r}{r_0}\right)^{2 \beta +\frac{4}{\beta -1}+7}}{4 (\beta -1) (\beta  (\beta +2)-1) r^3 (\chi +2 \pi ) (\chi +4 \pi )}.\label{m6}
\end{eqnarray}
\end{widetext}

Note that at the throat, one verifies 
\begin{eqnarray}
(\rho+ p)|_{r_0}= \frac{(\beta -1) \beta  r_0-(\beta +1) (2 \beta -1) (\beta  (\beta +2)-1) \tilde{C_1}}{\left(\beta  \left(\beta ^2+\beta -3\right)+1\right) r_0^3 (\chi +4 \pi )}. \nonumber \\
\end{eqnarray}

\begin{figure*}
    \includegraphics[scale=.6]{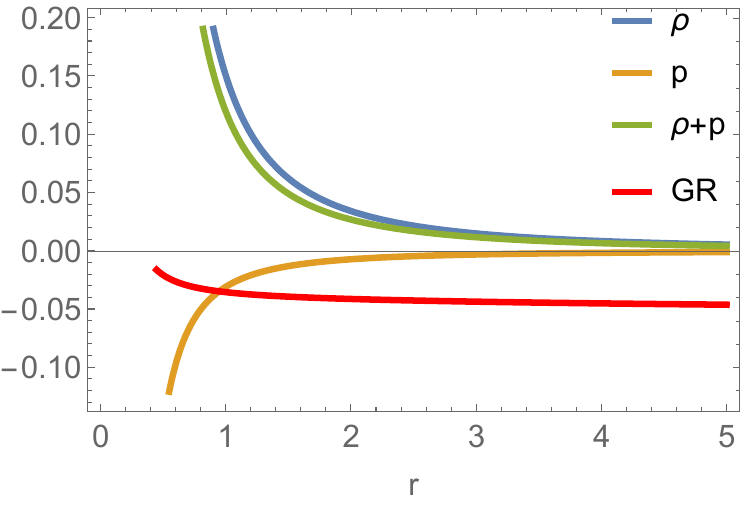}
    \includegraphics[scale=.6]{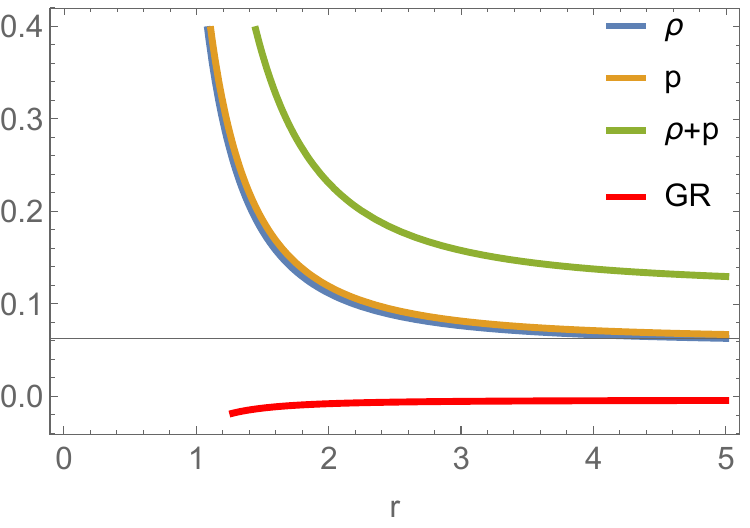} 
\caption{The figures show the energy density, pressure and NEC profile versus $r$. It is shown that the energy density is always positive throughout the spacetime, while the pressure negative when $\chi= 0.2$ and positive for $\chi= -10$. The standard NEC always holds with $\rho+p>0$. The constants are
$\beta=0.56, 0.6, \tilde{C_1}=0.05, 0.5$ and $r_0 =1$ respectively, for the left and right plots.  For the GR case we consider $\chi= 0$. }\label{f1}
\end{figure*}

The regions where the flaring-out condition $b'(r_0) <1$ is satisfied, we found the general condition $(\rho + p)|_{r_0}>0$, holds. Note that, if the parameter lies in the range $0.5<\beta< 1$, one may have wormholes in $f(R,T)$ gravity
satisfying the NEC at the throat.

Imposing the values of $\beta$, we plot the quantities $\rho$, $p$ and $\rho+p$ in Fig. \ref{f1} using the field Eqs. (\ref{m4})-(\ref{m6}). The plot shows that energy density $\rho >0$ is obeyed everywhere, including the wormhole throat; which means that the NEC is satisfied for all values of $r$. Nevertheless, the matter stress energy is seen to satisfy WEC inequalities also. For the Fig. \ref{f1}, we consider two sets of parameters (a) $\beta=0.56,  \tilde{C_1}=0.05,  r_0 =1$ and $\chi= 0.2$; (b) $\beta= 0.6, \tilde{C_1}= 0.5, r_0 =1$ and $\chi=  -10$, respectively. From our analysis it is clear that by choosing suitable values for the constants in order to have normal matter in the vicinity of the throat. This is important because wormholes are known to violate the energy conditions. Note that for $\chi=0$, one can obtain the general relativistic limit, where the matter violates the NEC at the vicinity of the throat
(see ed curve in Fig. \ref{f1}).

\subsection{Form function: $b(r)= r_{0}$}

It is well known that a simple class of solutions corresponds to $b(r)= r_{0}$. Thus, the
field Eq. (\ref{sp3}), provides the following redshift function as
\begin{widetext}
\begin{eqnarray}\label{d1}
\Phi (r)= C_2-\frac{1}{2} \log (r)+\log \left(4 C_1 \sqrt{r-r_0}+\sqrt{r} \left(2 r^2+5 r r_{0} -15 r_{0}^2\right)+15 r_{0}^2 \sqrt{r-r_{0}} \log \left(\sqrt{r-r_{0}}+\sqrt{r}\right)\right),
\end{eqnarray}
\end{widetext}
where $C_1$ and $C_2$ are constants of integration may be determine from the boundary conditions. We also verify that Eq. (\ref{d1}) diverges when $r \rightarrow \infty $ i.e., the redshift $\Phi(r)$ is not finite throughout the spacetime. However, according to the necessary criteria $\Phi(r)$ must be finite everywhere to avoid an event horizon.  However, one may avoid this situation by considering solutions with a cut-off of the stress-energy, see Refs. \cite{Lobo:2005vc,Lobo:2005yv} for more descriptive discussion. Following the standard procedure, one can also consider a flat space in the asymptotic limit by matching the interior solution to an exterior vacuum spacetime at a junction interface, $a$. Here, the obtained solution is not asymptotically flat, and we decide to restrict the solution at the throat neighbourhood so that the dimensions of these $f(R,T)$ wormholes are not arbitrarily large.

For simplicity, in this paper, we consider the exterior solution is the Schwarzschild spacetime,  so that the matching occurs at a radius greater than the event horizon $r_h = 2M$, i.e., $r = a> r_h $,  in order to avoid a black hole solution. Using the procedure as mentioned above, one may verify that the solutions become a wormhole, as the redshift function is finite in  the range $r_0 \leq r \leq a$ (see more details in  \cite{Banerjee:2019wjj}).

Thus, from Eqs. (\ref{sp1}) and  (\ref{sp2}), the
energy density and pressure are given by
\begin{widetext}
\begin{eqnarray}\label{d2}
\rho(r)=\frac{\sqrt{r} \chi }{(\chi +2 \pi ) (\chi +4 \pi ) \left(4 C_1 \sqrt{r-r_0}+\sqrt{r} \left(2 r^2+5 r r_0-15 r_0^2\right)+15 r_0^2 \sqrt{r-r_0} \log \left(\sqrt{r-r_0}+\sqrt{r}\right)\right)},\\ 
p(r)=\frac{\sqrt{r} (3 \chi +8 \pi )}{(\chi +2 \pi ) (\chi +4 \pi ) \left(4 C_1 \sqrt{r-r_0}+\sqrt{r} \left(2 r^2+5 r r_0-15 r_0^2\right)+15 r_0^2 \sqrt{r-r_0} \log \left(\sqrt{r-r_0}+\sqrt{r}\right)\right)},\label{d3}\\
\rho(r)+p(r)=\frac{4 \sqrt{r}}{(\chi +4 \pi ) \left(4 C_1 \sqrt{r-r_{0}}+\sqrt{r} \left(2 r^2+5 r r_{0}-15 r_{0}^2\right)+15 r_{0}^2 \sqrt{r-r_{0}} \log \left(\sqrt{r-r_{0}}+\sqrt{r}\right)\right)}.\label{d4}
\end{eqnarray}
\end{widetext}
Note that we always consider the case $r\geq r_0 $ for positive value within the square root. 
To be specific, the energy density $\rho$ is positive  in the range $-4\pi< \chi < -2 \pi$ and for $\chi > 0 $. 

\begin{figure*}
    \includegraphics[scale=.6]{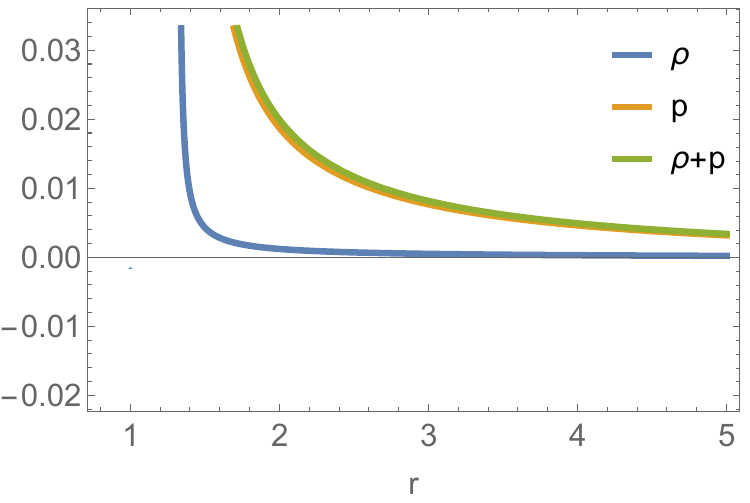}
    \includegraphics[scale=.6]{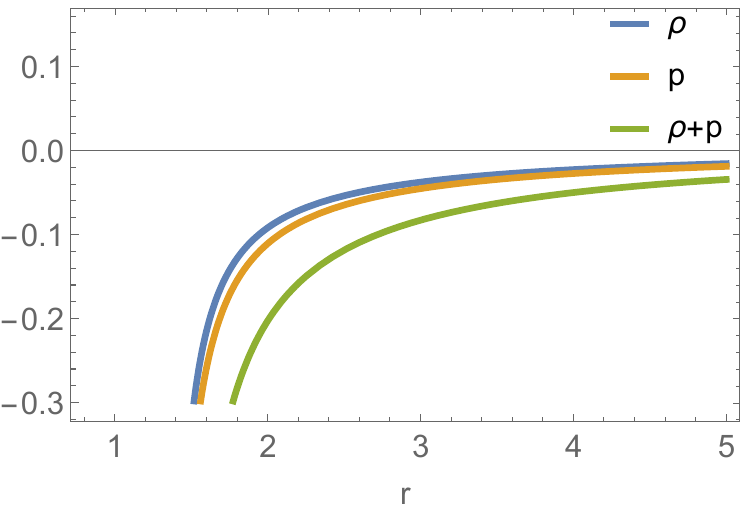}
\caption{The energy density, pressure and NEC profile for the specific case of  $b(r)=r_0$. In the left panel we found $\rho >0$ and $\rho+p >0$ outside the throat for $\chi>0$. Interestingly for $\chi>0$ the interior solution violates the NEC. This situation is same for $-4\pi< \chi < -2 \pi$. Relative to the NEC, $\rho+p >0$ at $r=r_0$ for $\chi< -4\pi $ but outside the throat $\rho+p <0$. Thus, the NEC is not satisfied everywhere. The values of the free parameters are $ {C_1}=2, \chi= 2, -13$ and $ r_0 =1$ respectively, for the left and right plots. }\label{f2}
\end{figure*}

The NEC at the throat is given by
\begin{eqnarray}\label{d5}
(\rho+p)\big|_{r_0}= -\frac{1}{2 r_0^2 (\chi +4 \pi )}.
\end{eqnarray}
The properties of the NEC changes at the throat. If we want to preserve the NEC at the throat, we have to set  $\chi < -4 \pi$ $\approx -12.6$, but the energy density is not positive in the radial region $r\geq r_0 $. This situation is very interesting in the sense that if $\chi > -4 \pi$, then the general condition $\rho+p < 0 $ at the throat. 

This restrictions are shown graphically in Fig. \ref{f2}  for the values of $\chi=2$ and $\chi= -13$. The Fig. \ref{f2}  shows that when $\chi > 0$, NEC is always satisfied outside the throat but violated for the normal matter threading the throat. This situation is same for $-4\pi< \chi < -2 \pi$. However, for
large negative values of $\chi=-13$, one verifies that the NEC is satisfied at the throat, but 
the value $\rho+p$ is negative in the radial region $r_0 \leq r \leq a$. In addition to this, in the context of classical relativity the energy density is zero for $b(r) = r_0$, whilst in $f(R,T)$ gravity it is positive for $ \chi < -4 \pi$.

It must be noted that it is of specific interest
to consider a thin shell with $r = a > r_0$, so that the dimensions of these wormholes are not arbitrarily large.

\section{Solution with specific transformation}\label{sec5}

In this section, we continue to work out an exact solution by introducing a new substitution $\Phi'(r)=\Psi(r)$, then  Eq. (\ref{sp3}) simplifies to
\begin{eqnarray}\label{s1}
\Psi '(r)=\frac{b' r-3b}{2r^2(r-b)}+\frac{b' r-3b+2b}{2r(r-b)}\Psi -\Psi^2 ,
\end{eqnarray}
It is now possible to write the Eq. (\ref{s1}) into a second order homogeneous equation with the unknown $\Psi (r)=\frac{u'(r)}{u(r)}$. Therefore, the general expression brought to \cite{Halder:2019urh} 
\begin{eqnarray}\label{s2}
\frac{d^2 u}{dr^2}-\frac{b' r-3b+2b}{2r(r-b)} \frac{d u}{dr}-\frac{b' r-3b}{2r^2(r-b)}=0.
\end{eqnarray}
With the Eq. (\ref{s2}), an additional restriction is necessary in order to close the system. For this purpose, we adopt the traditional approach, i.e.,
choosing the shape function and establish the existence of a  wormhole model. 

\begin{figure*}
    \includegraphics[scale=.6]{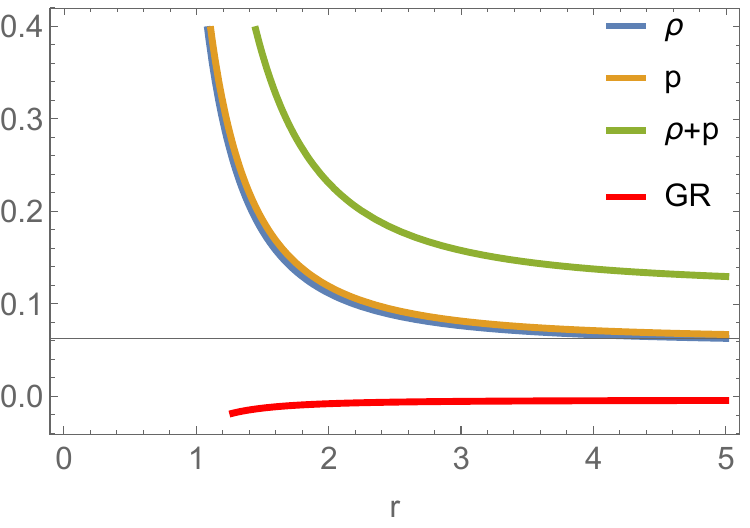}
    \includegraphics[scale=.6]{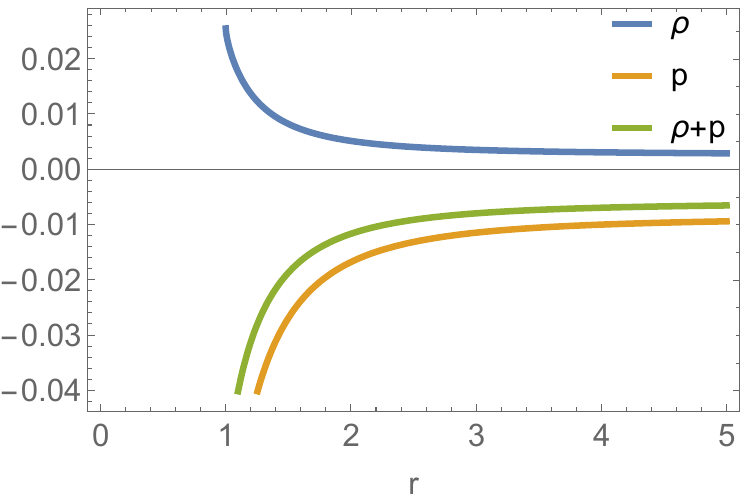}
\caption{The energy density, pressure and NEC profile for the transforming isotropic equation. The $\chi$ is restricted within the range $\chi< -4\pi$, with the other parameters $ C_1=0.5, C_2=-0.05, r_0 =1$ and $\chi= -13$. In the left panel we 
found $\rho >0$ and $\rho+p >0$. This wormhole solution obeys the matter NEC everywhere in conformity with our aim.
Other wormhole solutions are violating NEC when  $\chi> -4\pi$ with other specified parameters in the text.}\label{f3}
\end{figure*}
 
\subsubsection{Shape function $b(r)= \sqrt{r_0 r}$}
Let us fix t shape function $b(r)= \sqrt{r_0 r}$, then one finds the $u(r)$ as
 \begin{widetext}
 \begin{equation}
u(r)= \frac{c_2 \sqrt{r} \left(2 r^2+5 r \sqrt{\text{r}r_0}-15 \text{r}r_0\right)+4 c_1 \sqrt{r-\sqrt{\text{r}r_0}}+15 c_2 \text{r}r_0 \sqrt{r-\sqrt{\text{r}r_0}} \log \left(\sqrt{r-\sqrt{\text{r}r_0}}+\sqrt{r}\right)}{4 \sqrt{r}} .
\end{equation}
\end{widetext} 
 Without loss of generality one may determine
the following redshift function $e^{2\Phi}=Au^2(r)$. But the condition $e^{2\Phi} \rightarrow 0$ when $r \rightarrow \infty$, is not satisfied. So, the constants of integration $c_1$ and $c_2$ may be determined from the boundary conditions, $\Phi(a)$, at the junction interface. Consequently, the stress energy tensor components are given by
\begin{widetext}
\begin{eqnarray}\label{D46}
&&\rho(r)=\frac{\chi  \left(c_2 \Upsilon_1+c_2 r^{5/2} \left(15 r_0+2 \sqrt{\text{r}r_0}\right)+8 c_2 r^{7/2}+ c_2 \Upsilon_2 \sqrt{r-\sqrt{\text{r}r_0}}  +4 c_1 \sqrt{\text{r}r_0} \sqrt{r-\sqrt{\text{r}r_0}}-4 \sqrt{r} \sqrt{\text{r}r_0}\right)}{32 r^{7/2} (\chi +2 \pi ) (\chi +4 \pi )},\\
&& p(r)=\frac{(3 \chi +8 \pi ) \left( c_2   \sqrt{r-\sqrt{\text{r}r_0}} \left(\Upsilon_1 + \Upsilon_3\right) + c_2  \left(r_0-\sqrt{\text{r}r_0}\right) \Upsilon_2+4 c_1 r \sqrt{\text{r}r_0}-4 c_1 \text{r}r_0-4 \sqrt{r} \sqrt{\text{r}r_0} \sqrt{r-\sqrt{\text{r}r_0}}\right)}{32 r^{7/2} (\chi +2 \pi ) (\chi +4 \pi ) \sqrt{r-\sqrt{\text{r}r_0}}}, \nonumber \\ \label{D47} \\
&& \rho(r)+p(r)= \frac{c_2 \Upsilon_1+c_2 r^{5/2} \left(15 r_0+2 \sqrt{\text{r}r_0}\right)+8 c_2 r^{7/2}+ c_2 \Upsilon_2 \sqrt{r-\sqrt{\text{r}r_0}} +4 c_1 \sqrt{\text{r}r_0} \sqrt{r-\sqrt{\text{r}r_0}}-4 \sqrt{r} \sqrt{\text{r}r_0}}{8 r^{7/2} (\chi +4 \pi )},\label{D48}
\end{eqnarray}
\end{widetext}
 for notations simplicity we use
 $\Upsilon_1(r)= -5 r^{3/2} \left(3  r_0 \sqrt{\text{r}r_0}+2  \text{r}r_0\right)$,
$ \Upsilon_2(r)= 15r r_0 \left(2 r-\sqrt{\text{r}r_0}\right) \log \left(\sqrt{r-\sqrt{\text{r}r_0}}+\sqrt{r}\right)$,
$\Upsilon_3(r)=r^{5/2}  \left(15 r_0+2 \sqrt{\text{r}r_0}\right)+8  r^{7/2}$. The reduced NEC equations when evaluated at the throat is given by
\begin{eqnarray}\label{D49}
(\rho+p)\big|_{r_0}=-\frac{1 }{2 r_0^{2} (\chi +4 \pi )},
\end{eqnarray}
For concreteness, we will focus our attention on the study of energy conditions. Here,
we consider the specific value of $r_0=1$. Using the expansion (\ref{D49}) the violation of NEC, in the interval $\chi> -4\pi$. Consequently, the NEC is satisfied at the throat, when $\chi< -4\pi$.

This is confirmed graphically for the cases $\chi= -13$ (left panel) and $\chi= -4$ (right panel), as depicted in Fig. \ref{f3}, where we consider $r_0=1$. Considering the expressions (\ref{D46})-(\ref{D48}), it's become clear that the energy density, pressure and the NEC are positive everywhere imposes the restriction  $\chi< -4\pi$, and thereby satisfying the WEC also as depicted in Fig. \ref{f3}. However, it is shown that this solution has violating NEC when $\chi=0$ (as in the left panel of Fig. \ref{f3} with red curve) which is necessary in the context of classical relativity.
Then, in the right panel we show the results we obtained for $\chi= -4$, using the same values considered above, i.e.  $ C_1=0.5, C_2=-0.05$, and $r_0 =1$. The NEC is violated for the normal matter threading
the throat, as can be readily verified from Fig. \ref{f3}. In this context, one can also notice that it is possible to minimize the NEC violating matter for increasing values of $\chi$ in positive direction. 

\begin{figure*}
    \includegraphics[scale=.6]{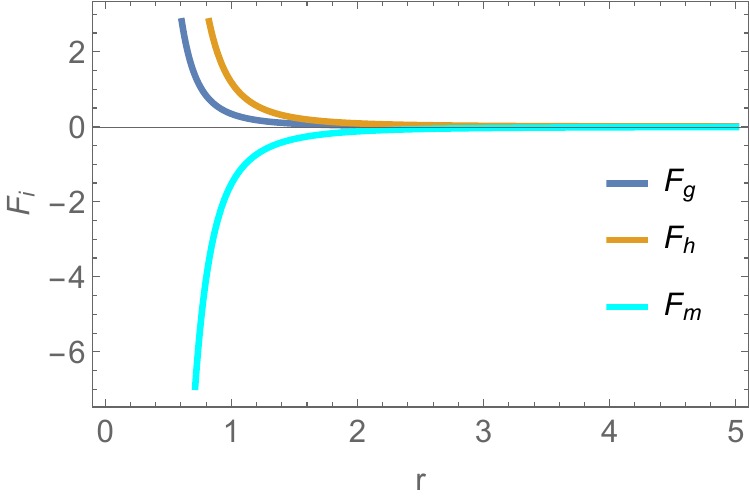}
    \includegraphics[scale=.62]{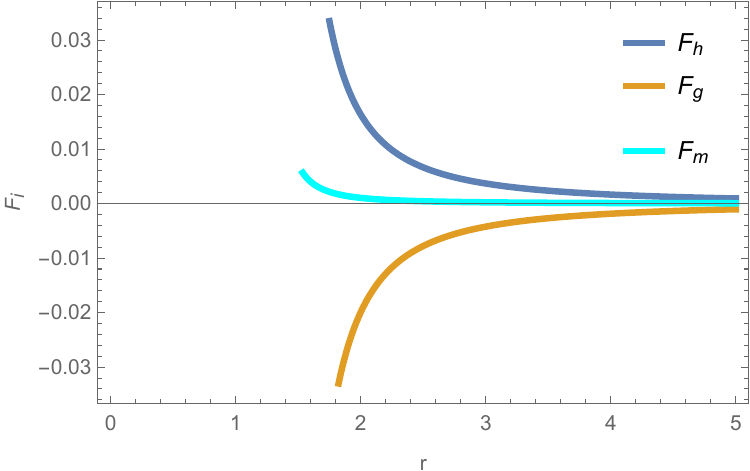}
\caption{Balancing behavior of $F_h$, $F_g$ and $F_m$ are plotted as a
function of the radial coordinate $r$ given in Eq. (\ref{eq45}). Here, we 
consider only those cases where NEC and WEC conditions are satisfied in the vicinity of the wormhole throat. The values of the parameters are $\chi= -10$, $\beta=0.56, 0.6, \tilde{C_1}=0.05, 0.5$ and $r_0 =1$ for left panel (case IV.2) and $ {C_1}=2, \chi= 2, -13$ and $ r_0 =1$ for right panel (case IV.3), respectively. }\label{f4}
\end{figure*}
\begin{figure}
    \includegraphics[scale=.62]{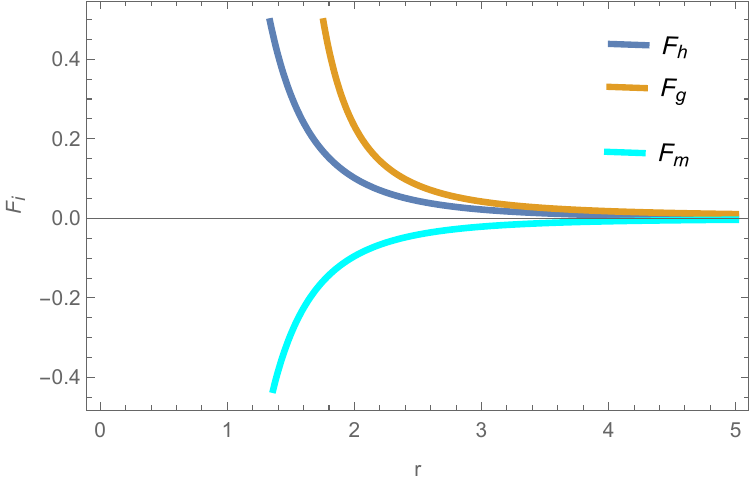}
\caption{Balancing behavior of different forces are plotted as a
function of the radial coordinate $r$. The values of the parameters are  
$ C_1=0.5, C_2=-0.05, r_0 =1$ and $\chi= -13$ for case V.}\label{f5}
\end{figure}

\subsubsection{Shape function $b(r) =r_0^2/r$}
For this case Eq. (\ref{s2}) gives
\begin{eqnarray}
u(r) = (-1)^{\frac{1}{4} \left(3-\sqrt{17}\right)} r^{\frac{3}{2}-\frac{\sqrt{17}}{2}} r_0^{-\frac{\sqrt{17}}{2}-\frac{3}{2}} \nonumber \\ 
\left(c_1 r_0^{\sqrt{17}}\xi_1 
+i^{\sqrt{17}} c_2 r^{\sqrt{17}} \xi_2\right), 
\end{eqnarray}
and substituting this into $\Psi (r)=\frac{u'(r)}{u(r)}$, we have 
\begin{eqnarray}
\Psi (r)=\frac{1}{2 r \left(c_1 r_0^{\sqrt{17}} \xi_1+i^{\sqrt{17}} c_2 r^{\sqrt{17}} \xi_2\right)}\times \nonumber \\
\Big[c_1 r_0^{\sqrt{17}} \left(4 \xi_1-\left(\sqrt{17}+1\right)  \xi_3\right) 
+e^{\frac{1}{2} i \sqrt{17} \pi } c_2 r^{\sqrt{17}} \times \nonumber \\ 
\left(4 \xi_2+\left(\sqrt{17}-1\right) \xi_4\right)\Big],
\end{eqnarray}
where $_2F_1$ is the hypergeometric function with \\
$\xi_1=\, _2F_1\left(\frac{1}{4} \left(-\sqrt{17}-1\right),\frac{1}{4} \left(3-\sqrt{17}\right);1-\frac{\sqrt{17}}{2};\frac{r^2}{r_0^2}\right)$,

$\xi_2=\, _2F_1\left(\frac{1}{4} \left(\sqrt{17}-1\right),\frac{1}{4} \left(\sqrt{17}+3\right);\frac{1}{2} \left(\sqrt{17}+2\right);\frac{r^2}{r_0^2}\right)$,

$\xi_3= \, _2F_1\left(\frac{1}{4} \left(3-\sqrt{17}\right),\frac{1}{4} \left(3-\sqrt{17}\right);1-\frac{\sqrt{17}}{2};\frac{r^2}{r_0^2}\right)$,

$\xi_4=\, _2F_1\left(\frac{1}{4} \left(\sqrt{17}+3\right),\frac{1}{4} \left(\sqrt{17}+3\right);\frac{1}{2} \left(\sqrt{17}+2\right);\frac{r^2}{r_0^2}\right)$.
which is a complex number, and therefore have no real solution. We have verified using Mathematica that $e^{\Phi(r)}$ does not go to a finite limit when $r\rightarrow \infty$.  Thus, one cannot find an exact general solution for the shape function $b(r)=r_0^2/r$, with an arbitrary $\chi$.

\section{Hydrostatc equilibrium}\label{sec6}
At this point let us consider the equilibrium stage that
can be achieved for matter threading the wormhole in $f(R,T)$ gravity.  One important and elegant method allowing their investigation is by formulating modified 
TOV equation (\ref{eq12}). Along these lines, a wide variety of astrophysical solutions (including wormholes/compact stars) have been studied (for a review, see, e.g., Refs. \cite{Rahaman:2013xoa,Rahaman:2014dpa,Jawad:2015uea,Deb:2017voy}. For the given spherically
symmetric space-time, the modified form of the TOV equation is given by 
\begin{eqnarray}
-\frac{dp}{dr}-\Phi'(\rho+p)+\frac{\chi}{(8\pi+2\chi)}(\rho'-p')=0.
\end{eqnarray}
The above equation can be symbolized as
\begin{eqnarray}\label{eq45}
F_h+F_g+F_m =0,
\end{eqnarray}
where the first term represents the hydrodynamic force ($F_h$), the second term is gravitational force ($F_g$), and the last term represents a force ($F_m$) which arises due to modification of the gravitational Lagrangian of the Einstein-Hilbert action. The 
 Eq. (\ref{eq45}) predicts that for the system to be in equilibrium, the sum of different forces for our system must be zero.
 
The profiles of $F_h$, $F_g$ and $F_m$ are shown in Fig. \ref{f4} and Fig. \ref{f5}, respectively. Here, we consider those cases where the NEC and WEC are satisfied.  Observing the Figs. \ref{f4} and \ref{f5}, we show that in all cases the equilibrium of forces is achieved and this supports the stability of the system. Thus, non-minimal curvature-matter coupling process an extra force which
leads to the existence of an equilibrium condition for matter distribution comprising the wormhole.

\section{SUMMARY AND DISCUSSION}\label{sec7}
Wormholes are nontrivial objects which connect two separate and distinct spacetime regions. In order to get an exact wormhole geometry, it requires exotic matter, which in GR demands the NEC violation. In this paper, we have found novel wormholes in the context of $f(R,T)$ gravity with isotropic matter respecting NEC. More specifically, we consider the minimal theories of type $f(R,T)= R+2\chi T$ \cite{Harko:2011kv}, which has attracted significant attention is a simple, but effective theory.  We mainly focused on sources with isotropic pressures. However, most of the wormhole solutions are anisotropic. Solutions are classified into two types: the GR and non-GR branches, depending on the existence or absence of $\chi$.

We summarize the obtained results as follows:

\begin{itemize}
\item For a zero-tidal-force wormhole with a perfect fluid source, we show that it is not possible to sustain a  wormhole solution. This indicates that it is impossible to obtain a solution in modified gravity and GR \cite{Cataldo:2016dxq}. 

\item With the specific case of $\Phi(r)$, we explore asymptotically flat wormhole solutions. We showed that NEC is satisfied at the throat by considering $\chi=0.2$ and $\chi= -10$. We have also seen that obtained solutions satisfy WEC as $\rho>0$ throughout the spacetime. 

\item In next, we concentrate our solution for a specific shape function $b(r)=r_0$. 
This choice demands satisfying the NEC at the throat, in the interval $\chi < -4 \pi$ which yields
violation of NEC outside the throat. Moreover, $\rho >0$ and $\rho+p >0$ are positive outside the throat
for $-4\pi< \chi < -2 \pi$ and $\chi > 0 $, whereas at the throat NEC is violated. Thus, the NEC is not satisfied everywhere.

\item As the next step, we transform the isotropic equation to a second-order homogeneous equation, and
it is possible to find wormhole solutions for the specific shape function $b(r)=\sqrt{r_0 r}$. 
For this case, we also studied the validity of the energy conditions and have the matter NEC
in the interior being obeyed for $\chi< -4\pi$. With these pacific values of parameters,
we see that WEC holds everywhere else for $\chi< -4\pi$. 

\item Finally, we discuss the stability of the system in details by performing a modified TOV equation for $f (R, T)$ gravity as given in Eq. (\ref{eq45}). This allows us to suppose that distinct forces have achieved the static equilibrium in all the considered cases. 

\end{itemize}

There are two main exciting conclusions. On the one hand, by choosing adequate parameters, it is possible to find
wormhole solutions without the need for exotic matter. On the other hand, wormhole solutions satisfy NEC around the throat but violated beyond it.

\end{document}